\documentclass[aps,prl,twocolumn,superscriptaddress,showpacs]{revtex4-1}
\usepackage{amssymb,amsmath,amsfonts,graphicx,epsf}
\usepackage{color}

\begin{document}

\title{Electron and ion energization in relativistic plasma turbulence}

\author{Vladimir Zhdankin}
\email{zhdankin@princeton.edu}
\thanks{Einstein fellow}
\affiliation{Department of Astrophysical Sciences, Princeton University, Peyton Hall, Princeton, NJ 08544}
\affiliation{JILA, University of Colorado and NIST, 440 UCB, Boulder, CO 80309}
\author{Dmitri A. Uzdensky}
\affiliation{Center for Integrated Plasma Studies, Physics Department, University of Colorado, 390 UCB, Boulder, CO 80309} 
\author{Gregory R. Werner}
\affiliation{Center for Integrated Plasma Studies, Physics Department, University of Colorado, 390 UCB, Boulder, CO 80309}
\author{Mitchell C. Begelman}
\affiliation{JILA, University of Colorado and NIST, 440 UCB, Boulder, CO 80309}
\affiliation{Department of Astrophysical and Planetary Sciences, 391 UCB, Boulder, CO 80309}
%\thanks{}

\date{\today}

\begin{abstract}
Electron and ion energization (i.e., heating and nonthermal acceleration) is a fundamental, but poorly understood, outcome of plasma turbulence. In this work, we present new results on this topic from particle-in-cell simulations of driven turbulence in collisionless, relativistic electron-ion plasma. We focus on temperatures such that ions (protons) are sub-relativistic and electrons are ultra-relativistic, a regime relevant for high-energy astrophysical systems such as hot accretion flows onto black holes. We find that ions tend to be preferentially heated, gaining up to an order of magnitude more energy than electrons, and propose a simple empirical formula to describe the electron-ion energy partition as a function of the ratio of electron-to-ion gyroradii (which in turn is a function of initial temperatures and plasma beta). We also find that while efficient nonthermal particle acceleration occurs for both species in the ultra-relativistic regime, nonthermal electron populations are diminished with decreasing temperature whereas nonthermal ion populations are essentially unchanged. These results have implications for modeling and interpreting observations of hot accretion flows.
\end{abstract}


\maketitle

{\em Introduction.}---  Plasma energization via turbulent dissipation is a fundamental topic in plasma physics. It involves a number of important questions that are difficult to address with analytic theory, including: What fraction of injected energy is dissipated into each of the constituent particle species (electrons, ions, etc.)? Are there collisionless mechanisms of thermal coupling between electrons and ions, sufficient to keep their corresponding temperatures comparable? Is there nonthermal particle acceleration (NTPA), and if so, what are the characteristics of the resulting energetic electron and ion populations?

The answers to these questions have important implications for myriad space and astrophysical systems. For motivation in this paper, we consider the example of radiatively inefficient accretion flows (RIAFs) around black holes. RIAFs comprise tenuous, collisionless plasma with relativistic electrons subject to radiative cooling. Such an accretion flow is at risk of collapsing into a collisional thin disk. To explain the survival of~RIAFs, models require the ambient turbulence to heat ions preferentially, which establishes a ``two-temperature'' plasma with sufficient ion pressure to avoid collapse \citep{shapiro_lightman_eardley_1976, ichimaru_1977, rees_etal_1982, narayan_yi_1995, quataert_gruzinov_1999}. An accurate, comprehensive theoretical prescription for the electron and ion heating rates would be extremely valuable for phenomenological models or global magnetohydrodynamic (MHD) simulations of RIAFs \cite{ressler_etal_2015, foucart_etal_2015, ball_etal_2018b, ryan_etal_2018}. However, so far, only simplified (linearized, non-radiative, non-relativistic) analytic models \cite{howes_2010, quataert_gruzinov_1999} and empirical fitting formulae from idealized kinetic simulations \cite[e.g.,][]{matthaeus_etal_2016, rowan_sironi_narayan_2017, werner_etal_2018, kawazura_etal_2018} exist. Furthermore, the conceivable existence of collective plasma phenomena that thermally couple electrons and ions could complicate the establishment of high ion temperatures \cite{begelman_chiueh_1988, sironi_narayan_2015}.

RIAFs are also notable for their highly nonthermal radiative signatures \cite[e.g.,][]{yuan_etal_2003, remillard_mcclintock_2006, yuan_narayan_2014}. Understanding the physical processes responsible for the underlying NTPA is essential for interpreting observations. Collisionless plasma turbulence driven by the magnetorotational instability \cite{balbus_hawley_1991} is a primary candidate source of NTPA, with supporting evidence from local (shearing-box) kinetic simulations \cite[e.g.,][]{riquelme_etal_2012, hoshino_2013, hoshino_2015, kunz_etal_2016, inchingolo_etal_2018} and MHD test-particle simulations \cite{kimura_etal_2016}. An essential next step is to systematically determine the properties of NTPA in realistic parameter regimes at large system size.

First-principles kinetic simulations offer empirical insights necessary to build a rigorous understanding of electron and ion heating, thermal coupling, and NTPA in parameter regimes relevant for RIAFs. In this work, we use particle-in-cell (PIC) simulations of driven turbulence to study electron and ion energization (i.e., heating and NTPA) in relativistic plasmas. We focus on the regime where ions (protons) are sub-relativistic and electrons are ultra-relativistic, taking temperatures in the range $m_e c^2 \lesssim T \lesssim m_i c^2$, which we refer to as the {\it semirelativistic} regime \cite[][]{werner_etal_2018}. This regime is amenable to fully kinetic simulations using the real electron-proton mass ratio, as demonstrated by recent PIC studies of magnetic reconnection \cite{guo_etal_2016, rowan_sironi_narayan_2017, werner_etal_2018, ball_etal_2018}, due to the large relativistic mass of electrons reducing the kinetic scale separation with ions. Our results indicate that turbulence in this physical regime can efficiently energize ions, while electron energization becomes less efficient with decreasing temperature.

{~}\\
{\em Method.}--- The simulation set-up is similar to our previous work on pair (electron-positron) plasma turbulence \citep[e.g.,][]{zhdankin_etal_2018a}. We perform the simulations with the explicit electromagnetic PIC code {\sc Zeltron} \citep{cerutti_etal_2013} using charge-conserving current deposition \citep{esirkepov_2001}. The domain is a periodic cubic box of size $L^3$ with uniform mean magnetic field $\boldsymbol{B}_0 = B_0 \hat{\boldsymbol{z}}$. We initialize particles from a uniform Maxwell-J\"{u}ttner distribution with particle density per species $n_0$ and equal electron and ion temperatures, $T_e = T_i = T_0$. We then drive strong turbulence (with rms fluctuating magnetic field $\delta B_{\rm rms} \sim B_0$) at low wavenumber modes ($k = 2\pi/L$) by applying a randomly fluctuating external current density \citep{tenbarge_etal_2014}. We set $32$ particles per cell per species in all production runs.

There are then three free dimensionless parameters: the initial temperature relative to ion rest mass energy, $\theta_{i0} = T_0/m_i c^2$, the initial plasma beta,~$\beta_0 = 16 \pi n_0 T_0 / B_0^2$, and the ratio of the driving scale to the ion Larmor radius, $L/2\pi\rho_{i0}$ (subscript zero refers to initial values of parameters). The characteristic Larmor radii are given by $\rho_{s} = (\gamma_{s}^2 - 1)^{1/2} m_s c^2/e B_{\rm rms}$, where $\gamma_{s} = 1 + E_s/m_s c^2$ (for species $s \in \{e, i\}$) are the mean particle Lorentz factors, $E_s$ are the mean particle kinetic energies (for species $s$), and $B_{\rm rms}$ is the rms total magnetic field. In the semirelativistic regime ($m_e/m_i \ll \theta_{i0} \ll 1$), the separation between the electron and ion Larmor radii is given by $\rho_{e0}/\rho_{i0} \sim \theta_{i0}^{1/2}$, and the separation between the Larmor radius and skin depth scales as $\rho_{e0}/d_{e0} \sim \rho_{i0}/d_{i0} \sim \beta_0^{1/2}$. In the fully relativistic limit ($\theta_{i0} \gg 1$), the particle inertia is set by the relativistic mass, making the system similar to a pair plasma ($\rho_{e0} = \rho_{i0}$ and $d_{e0} = d_{i0}$). Thus, $\theta_{i0}$ controls the electron-ion scale separation. In our simulations, we fix the cell size to $\delta x = \min{(\rho_e/2, d_e/2)}$, i.e., relative to electron scales. For given plasma parameters, $L$ is thus proportional to the number of cells in each direction. For a fixed number of cells, obtaining a large ion kinetic range ($\rho_i/\rho_e$) comes at the expense of the inertial range ($L/2\pi\rho_i$), and vice versa. Finally, we note that in the semirelativistic regime, the initial Alfv\'{e}n velocity scales as $v_{A0}/c \sim (\theta_{i0}/\beta_0)^{1/2}$; thus, the turbulent motions become increasingly sub-relativistic with decreasing $\theta_{i0}$.

Our primary scan is performed with $256^3$-cell and $512^3$-cell simulations with $\theta_{i0}$ varying in the range $[1/2048, 10]$ at fixed $\beta_0 = 4/3$; we do a secondary scan with $\beta_0$ varying in the range $[1/12, 64/3]$ at fixed $\theta_{i0} = 1/16$. In addition, we performed three $768^3$ simulations with $\theta_{i0} \in \{ 1/1024, 1/256, 1/64 \}$ (at $\beta_0 = 4/3$) and one $1024^3$ simulation with $\theta_{i0} = 1/256$ and $\beta_0 = 4/3$ (and $L/2\pi\rho_{i0} = 8.8$ and $\rho_{i0}/\rho_{e0} = 9.3$).

\begin{figure}
\includegraphics[width=0.95\columnwidth]{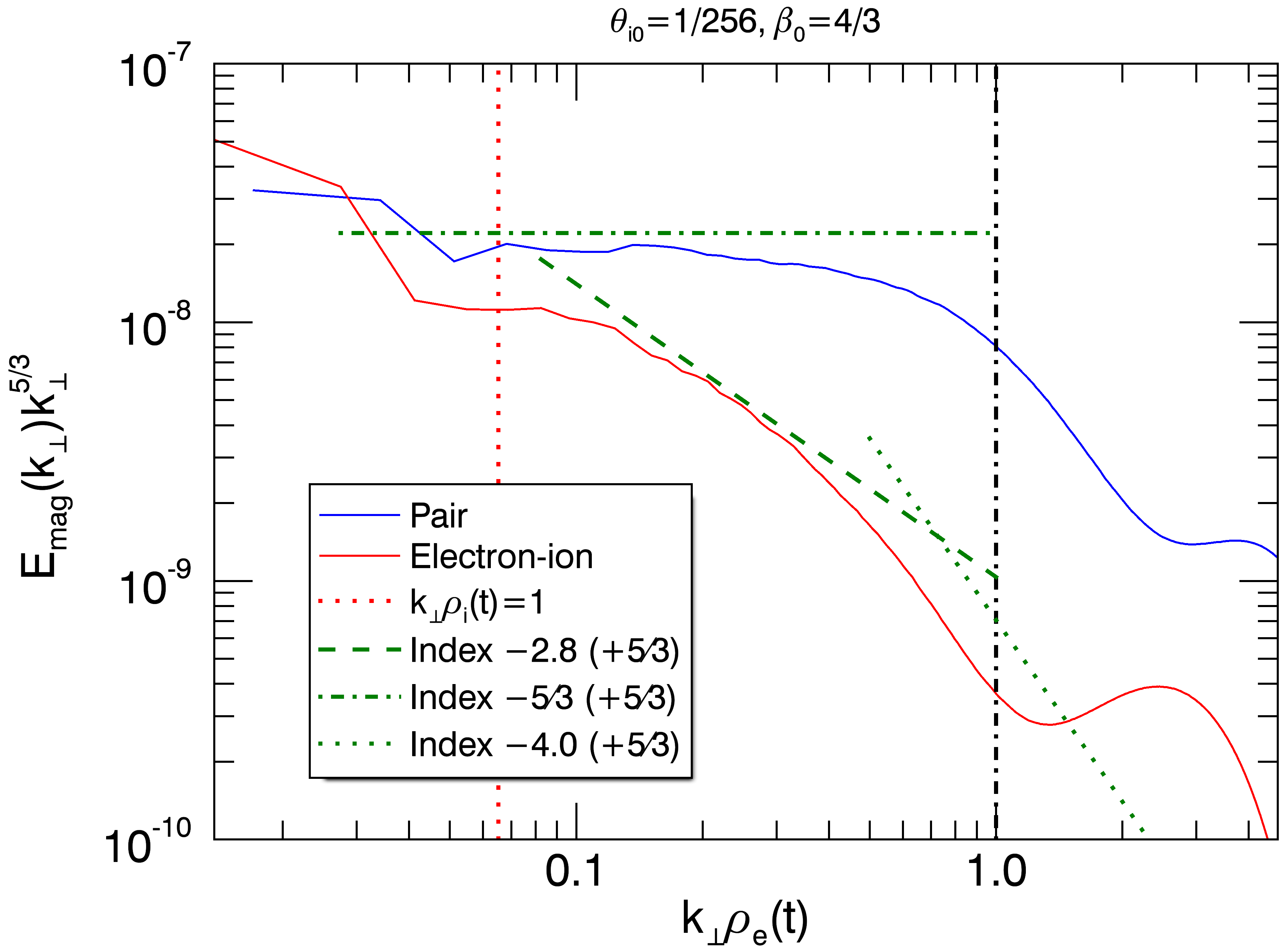}
\includegraphics[width=0.95\columnwidth]{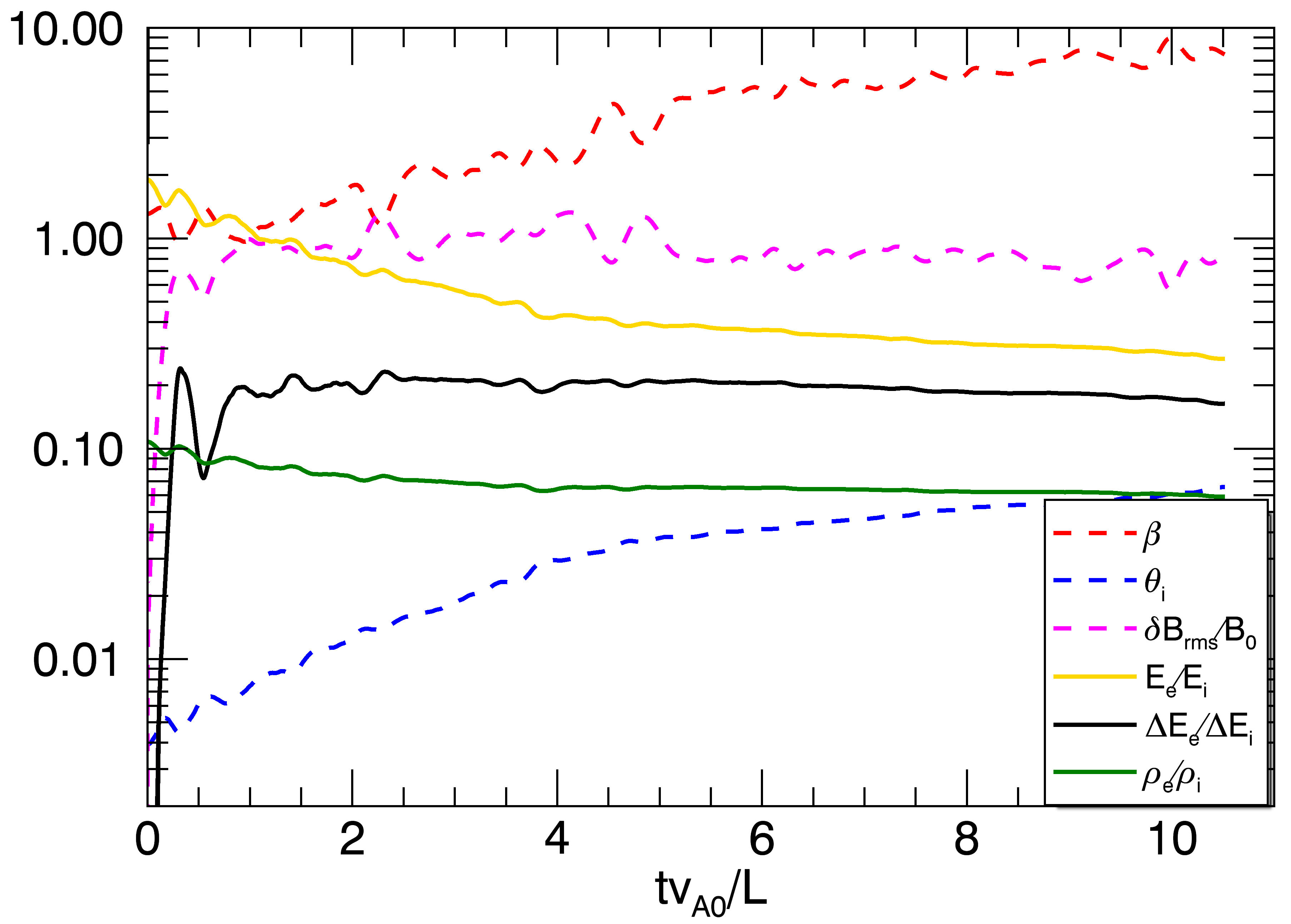}
   \centering
   \caption{\label{fig:spec} Top: Magnetic energy spectrum, compensated by $k_\perp^{5/3}$, for $1024^3$ electron-ion simulation (red) compared to a similar $1024^3$ pair-plasma simulation (blue). Power-law scalings are shown for reference (green). Bottom: Evolution of plasma parameters in the $1024^3$ simulation, including $\delta B_{\rm rms}/B_0$ (magenta), $\beta$ (red), and $\theta_i$ (blue; computed from particle energy assuming a thermal distribution). Also shown is the evolution of the electron-ion energy ratio $E_e/E_i$ (yellow), the electron-ion energy gain ratio $\Delta E_e/\Delta E_i$ (black), and ratio of electron-to-ion Larmor radii $\rho_e/\rho_i$ (green).}
 \end{figure} 

 \begin{figure}
\includegraphics[width=0.95\columnwidth]{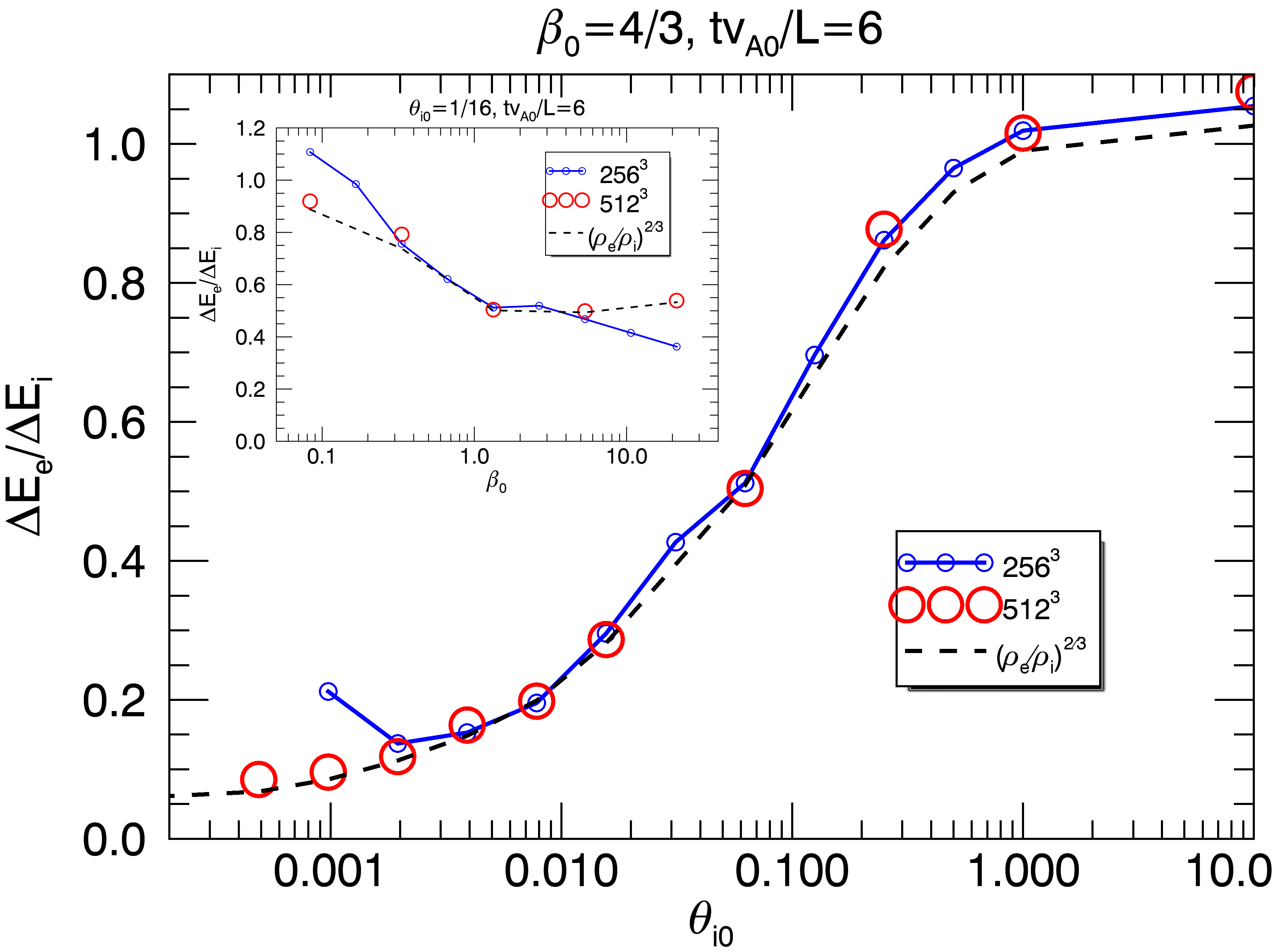}
\includegraphics[width=0.95\columnwidth]{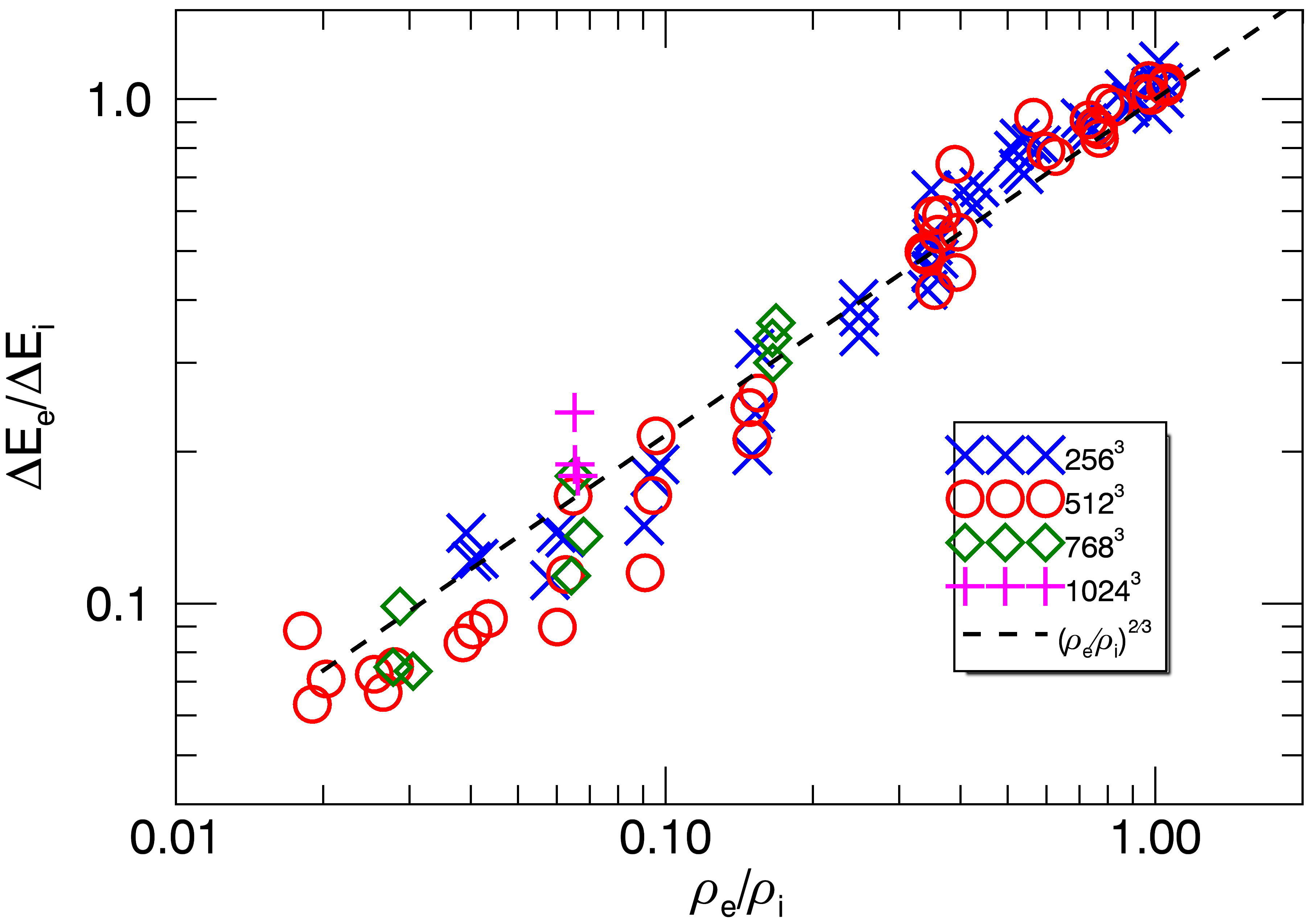}
      \centering
   \caption{\label{fig:partition} Top: Ratio of electron-to-ion energy gain, $\Delta E_e/\Delta E_i$, for varying $\theta_{i0}$ at $t v_{A0}/L = 6$ and $\beta_{0} = 4/3$; two sizes are compared, $256^3$ (blue) and $512^3$ (red). Inset: similar for varying $\beta_{0}$ (at fixed $\theta_{i0} = 1/16$). The fit by $(\rho_e/\rho_i)^{2/3}$ (measured at $t v_{A0}/L = 6$) is also shown (dashed). Bottom: $\Delta E_e/\Delta E_i$ versus mean $\rho_e/\rho_i$ for all simulations (ignoring $256^3$ cases that have not converged with system size), measured over intervals of duration $L/v_{A0}$ starting at $t v_{A0}/L \in \{ 3,4,5 \}$.}
 \end{figure}

{~}\\
{\em Results.}--- We first consider the magnetic energy spectrum, integrated over wavenumbers parallel to the guide field $\boldsymbol{B}_0$ and directions perpendicular to $\boldsymbol{B}_0$, which we denote $E_{\rm mag}(k_\perp)$, where $k_\perp$ is the wavenumber perpendicular to $\boldsymbol{B}_0$. We show $E_{\rm mag}(k_\perp)$ compensated by $k_\perp^{5/3}$ and time-averaged from $4.3 L/v_{A0}$ to $5.7 L/v_{A0}$, for the $1024^3$ case ($\theta_{i0} = 1/256$, $\beta_0 = 4/3$) in the top panel of Fig.~\ref{fig:spec}. To illustrate the effects of ions, we compare this to the spectrum from a similar $1024^3$ relativistic pair-plasma simulation (taken from our previous work \citep{zhdankin_etal_2018b}). Both simulations are consistent with a power-law spectrum with index near $-5/3$ at large scales ($k_\perp \rho_i \lesssim 1$ for electron-ion and $k_\perp \rho_e \lesssim 1$ for pair), broadly consistent with inertial-range MHD turbulence phenomenology \citep[e.g.,][]{goldreich_sridhar_1995}. The spectrum for the electron-ion case is significantly steeper in the ion kinetic range (between $k_\perp \rho_i =1$ and $k_\perp \rho_e = 1$), although not a clear power law; for reference, we show a comparison to a power law with index $-2.8$, often measured in the ion kinetic range for non-relativistic plasmas (including the solar wind \cite[e.g.,][]{alexandrova_etal_2009, sahraoui_etal_2009, kiyani_etal_2009, alexandrova_etal_2012, kiyani_etal_2015} and simulations \cite[e.g.][]{boldyrev_perez_2012, told_etal_2015, cerri_etal_2017, groselj_etal_2018}). A definitive measurement of the spectrum in the ion kinetic range requires larger simulations with lower $\theta_{i0}$ (larger $\rho_i/\rho_e$), in order to simultaneously resolve a long inertial range and ion kinetic range. In the electron kinetic range ($k_\perp \rho_e \gtrsim 1$), there appears to be a power law with index near $-4$, similar to the sub-Larmor spectrum in the pair-plasma case \citep{zhdankin_etal_2018a}.
 
 In the bottom panel of Fig.~\ref{fig:spec}, we show the evolution of the physical parameters $\beta$, $\delta B_{\rm rms}/B_0$, and $\theta_i$ (defined as $2 E_i/ 3 m_i c^2$) for the $1024^3$ simulation. Due to continuous energy injection, $\theta_i$ and $\beta$ both steadily increase over the simulation. We also show the electron-to-ion ratios of kinetic energy gains $\Delta E_e/\Delta E_i$, total kinetic energies $E_e/E_i$, and Larmor radii $\rho_e/\rho_i$. We find that these measured quantities vary only weakly with time after $\sim 3 L/v_{A0}$; this is also true in most of our other simulations (not shown). The approximate constancy of these quantities with time allows us to perform robust measurements of energy partition, as we discuss next.

In Fig.~\ref{fig:partition}, we show the ratio of electron-to-ion energy gain, $\Delta E_e/\Delta E_i$, as a function of various parameters. In the top panel and inset, we show $\Delta E_e/\Delta E_i$ as a function of $\theta_{i0}$ (at fixed $\beta_0 = 4/3$) and as a function of $\beta_0$ (at fixed $\theta_{i0} = 1/16$), measured from the initial time to $t v_{A0}/L = 6$ (arbitrarily chosen; other times give similar results). We find that $\Delta E_e/\Delta E_i \approx 1$ at $\theta_{i0} \gtrsim 1$ (i.e., in the relativistic regime, as expected theoretically) and decreases with lower $\theta_{i0}$, reaching $\Delta E_e/\Delta E_i \sim 0.1$ at $\theta_{i0} = 1/2048$ (near the transition between the semirelativistic regime and the fully sub-relativistic regime, where $T_0/m_e c^2 = \theta_{i0} m_i/m_e \sim 1$). Comparison of the $256^3$ and $512^3$ simulations indicates that results are converged with respect to system size except at sufficiently low $\theta_{i0}$, low $\beta_0$, or high $\beta_0$ (in these exceptions, $L/2\pi\rho_i \sim 1$ at late times so the driving interferes with kinetic processes).
 
 \begin{figure}
\includegraphics[width=0.95\columnwidth]{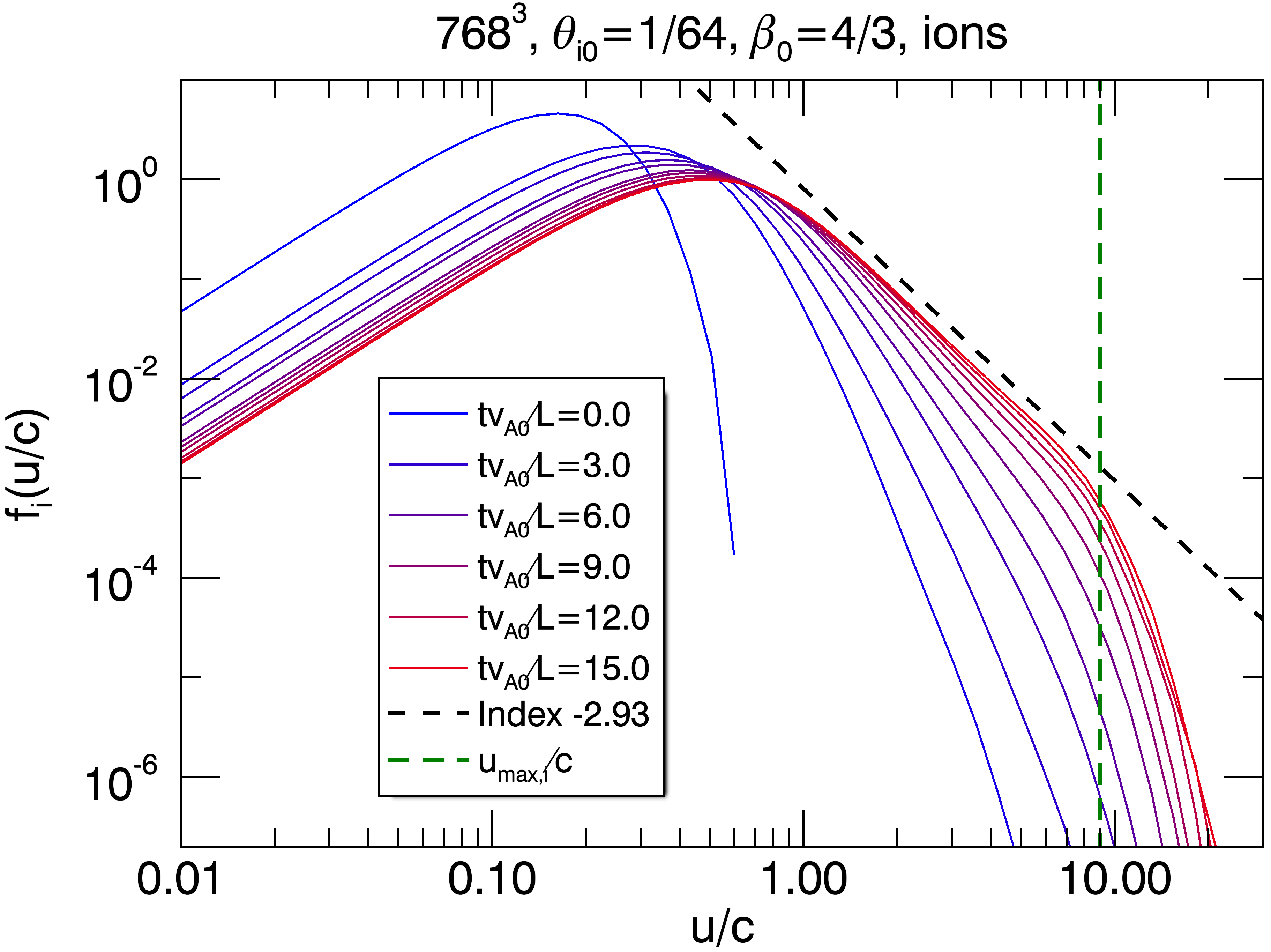}
\includegraphics[width=0.95\columnwidth]{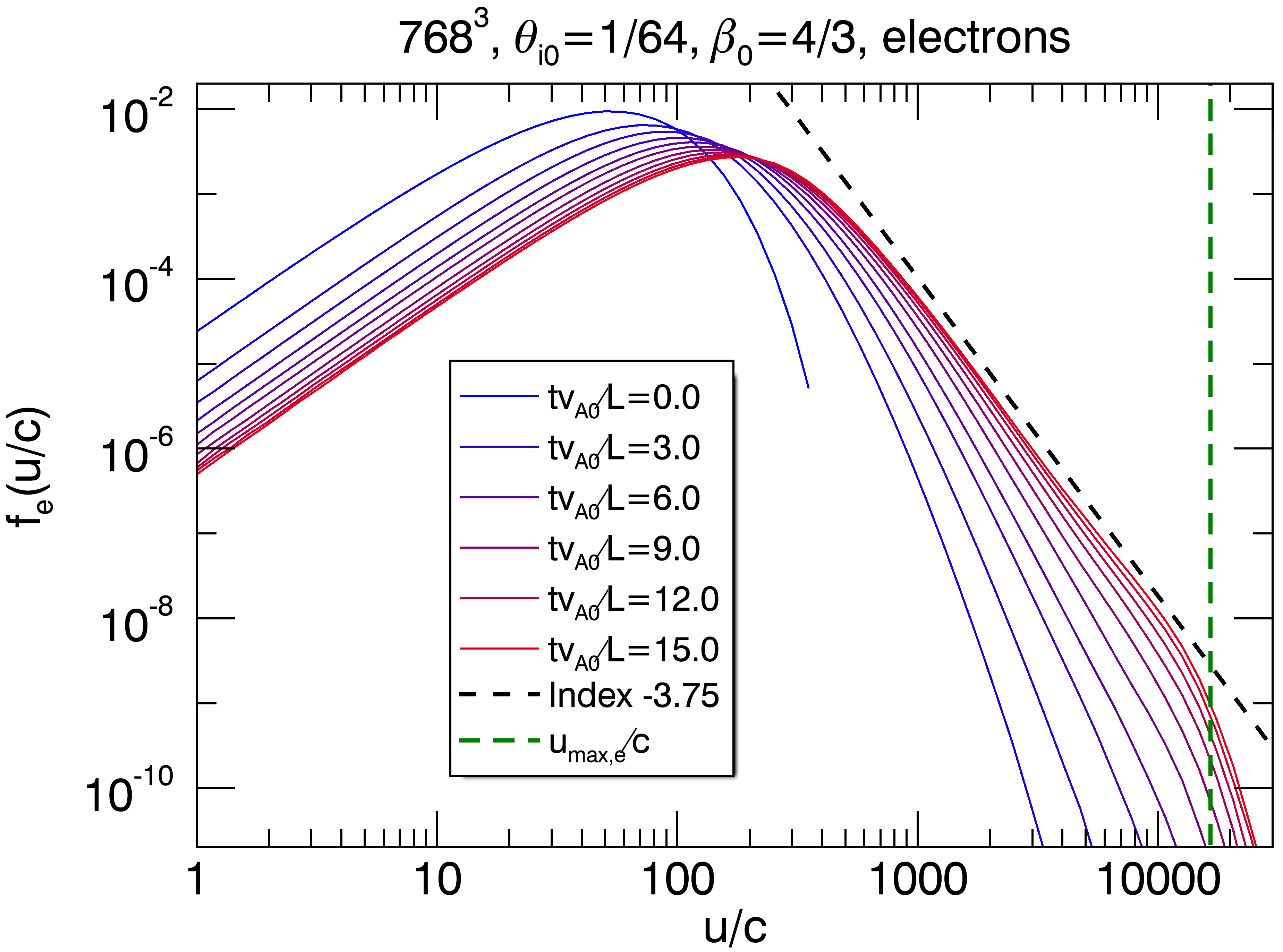}
\includegraphics[width=0.95\columnwidth]{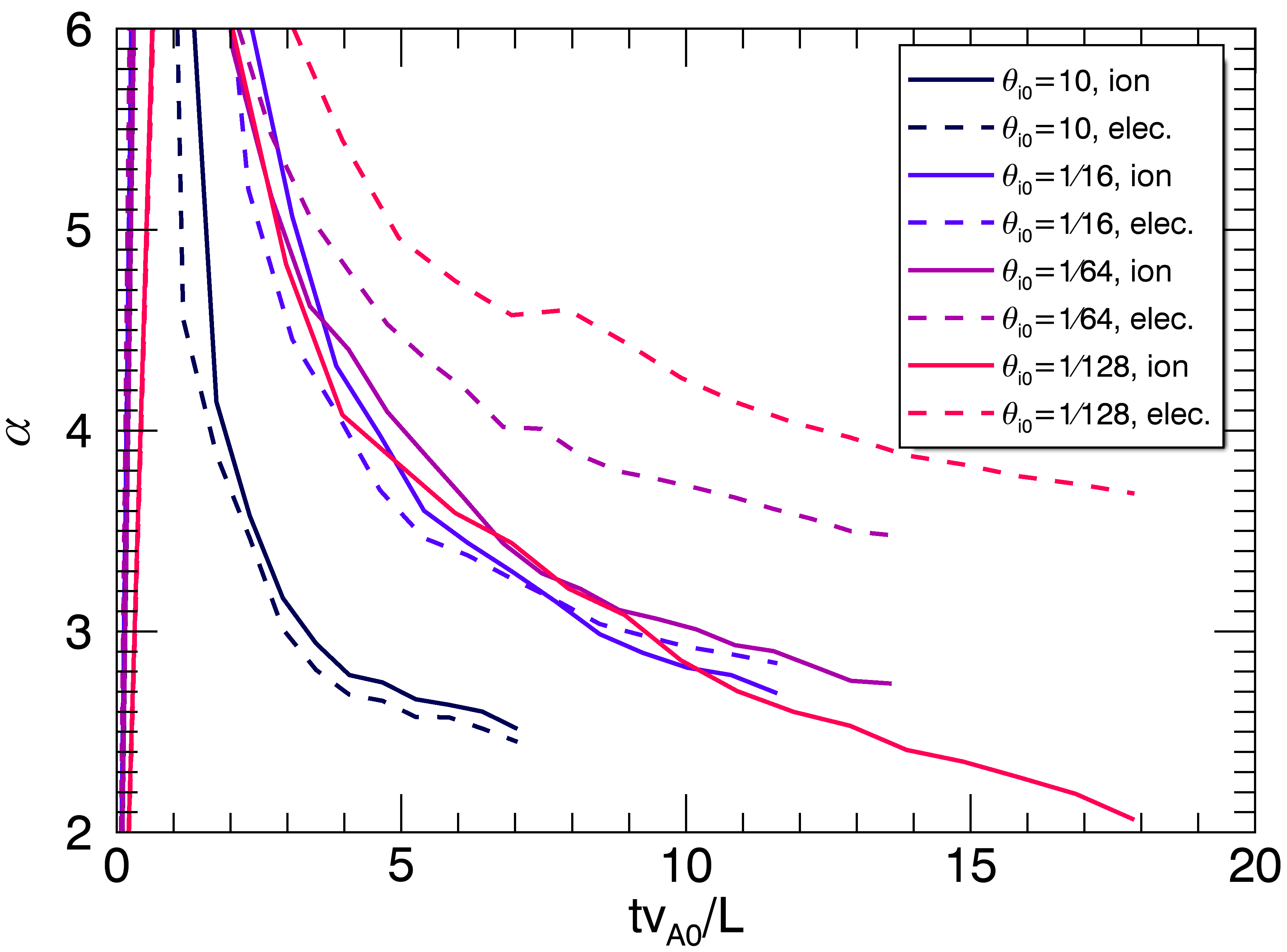}
   \centering
   \caption{\label{fig:distributions} Top: evolution of ion four-velocity distribution $f_i(u/c)$ to a power law with fitted index $\alpha_i \approx 2.9$ (black dashed), spanning up to $u_{{\rm max},i}$ (green dashed). Middle: similar for electron four-velocity distribution $f_e(u/c)$, which evolves to a power law with fitted index $\alpha_e \approx 3.8$. Bottom: evolution of the fitted power-law indices $\alpha_s$ for ions (solid) and electrons (dashed) for $512^3$ simulations with varying $\theta_{i0}$.}
 \end{figure}
 
Intriguingly, we find that the results can be well fit by the time-dependent empirical formula,
\begin{align}
\Delta E_e/ \Delta E_i \sim (\rho_e/\rho_i)^{2/3} \, . \label{eq:emp}
\end{align}
Note that $\rho_e/\rho_i$ is a nontrivial function of $\theta_{i0}$, $\beta_0$, and, to a lesser extent, time. The bottom panel of Fig.~\ref{fig:partition} explicitly shows the scaling of Eq.~\ref{eq:emp} compared to all simulations in our scan, where $\Delta E_e/\Delta E_i$ is now measured over intervals of duration $L/v_{A0}$ starting at $t v_{A0}/L \in \{ 3, 4, 5 \}$, to represent the short-term heating during fully developed turbulence.
 
 We next describe results on NTPA. In Fig.~\ref{fig:distributions}, we show the time evolution of the distributions $f_s(u)$ of four-velocities $u$, for ions and electrons ($s \in \{e, i\}$) in a representative semirelativistic case ($768^3$, $\theta_{i0} = 1/64$, $\beta_0 = 4/3$). We find that power-law tails gradually form over a number of dynamical times ($\sim 15 L/v_{A0}$, in this case), and become fully developed when the most energetic particles begin to accumulate at the system-size limited velocity, $u_{{\rm max},s} = L e B_0/2 m_s c$. To characterize the distributions, we measure the power-law indices $-\alpha_s = d\log{f_s}/d\log{u}$ at the geometric mean of the peak of the distribution and $u_{{\rm max},s}$. The ion distribution attains a fitted power-law index $\alpha_i \approx 2.9$, while the electrons attain $\alpha_e \approx 3.8$, indicating that ion acceleration is more efficient in this regime. Intriguingly, the late-time power-law index for ions is similar to that for the relativistic pair-plasma case at the same plasma beta \cite{zhdankin_etal_2018b}. Note that nonthermal ions are essentially relativistic in this example, despite being initialized well within the sub-relativistic regime; studying the transition of the power law through $u/c \sim 1$ will require even larger simulations with lower $\theta_{i0}$.
 
To illustrate the parameter dependence of the nonthermal distributions, we show the time evolution of $\alpha_s$ for simulations with varying $\theta_{i0}$ (fixed $\beta_0 = 4/3$) in the bottom panel of Fig.~\ref{fig:distributions}. Note that $\alpha_s$ decreases in time and does not saturate at a well-defined value, due to the pile-up of particles near $u_{{\rm max},s}$ influencing the measurement of $\alpha_s$ at late times \cite[c.f.,][]{zhdankin_etal_2018b}. We find that ions always reach $\alpha_i \sim 3$ before the pile-up becomes significant; this is a similar value to that in the fully relativistic case ($\theta_{i0} = 10$), although it takes a longer time to reach this value at low $\theta_{i0}$ (consistent with the diffusive particle acceleration timescale increasing with decreasing $v_{A0}/c$ \cite{zhdankin_etal_2018b}). The electron distributions, however, become softer (i.e., larger $\alpha_e$) when $\theta_{i0}$ is decreased. Hence, our results indicate that NTPA for ions remains as efficient in the semirelativistic regime as in the ultrarelativistic regime, while it becomes inefficient for electrons in the limit of low $\theta_{i0}$.
 
To further characterize the NTPA, we decompose the particle distributions into thermal and nonthermal components. To do this, we define the thermal part to be a Maxwell-J\"{u}ttner distribution with temperature and normalization such that the corresponding peak coincides with the peak of the measured distribution; we also consider any excess of the measured distribution at energies below the peak value to be part of the thermal component. The nonthermal population is then defined to be the difference between the measured distribution and the thermal fit.
 
\begin{figure}
\includegraphics[width=0.95\columnwidth]{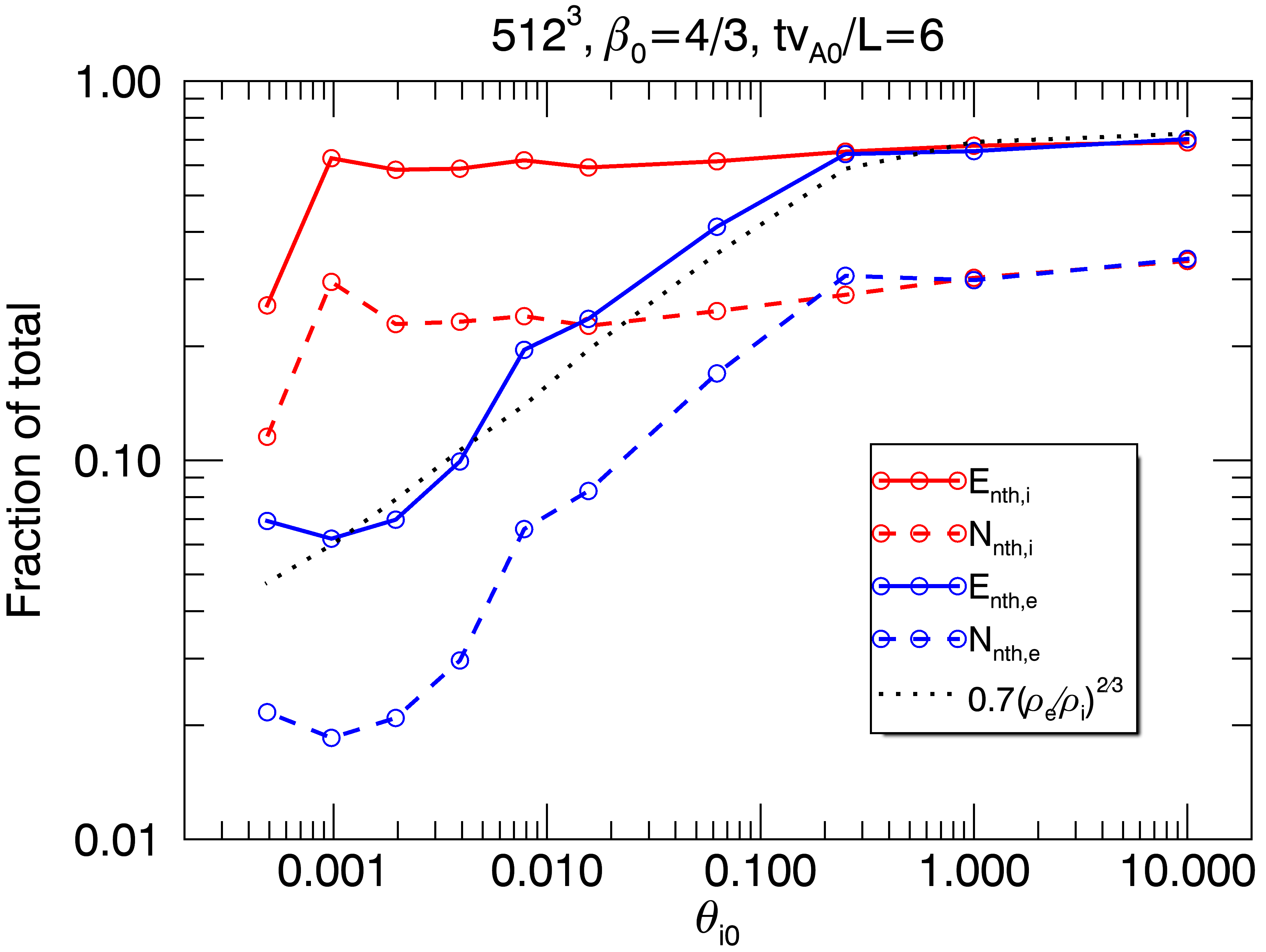}
      \centering
   \caption{\label{fig:decomp} Nonthermal energy (solid) and number (dashed) fractions for ions (red) and electrons (blue). The scaling $0.7 (\rho_e/\rho_i)^{2/3}$, tracing energy partition, is also shown (black).}
 \end{figure}

We show the fraction of the kinetic energy in the nonthermal population, $E_{{\rm nth},s}$, and the fraction of particles in the nonthermal population, $N_{{\rm nth},s}$, for electrons and ions ($s\in\{e,i\}$) as functions of $\theta_{i0}$ (fixed $\beta_0 = 4/3$ and $t v_{A0}/L = 6$) in Fig.~\ref{fig:decomp}. We find that the nonthermal energy fraction is roughly constant for ions (between $60-70\%$), but declines with decreasing $\theta_{i0}$ for electrons (from $\sim 70\%$ to $\sim 6\%$). The nonthermal number fraction is qualitatively similar to this, with $\sim 30\%$ of both particle species being nonthermal for $\theta_{i0} \gtrsim 1$ and the fraction of nonthermal electrons declining to $\sim 2\%$ at low $\theta_{i0}$. This confirms that nonthermal ion energization is significant in the semirelativistic regime, while electron energization is diminished. Interestingly, we find that the dependence of $E_{{\rm nth},e}$ on $\theta_{i0}$ mirrors the overall energy partition, being well fit by $0.7 (\rho_e/\rho_i)^{2/3}$ (similar to Eq.~\ref{eq:emp}), suggesting that NTPA is linked to the available energy budget. The nonthermal fractions also have a $\beta_0$ dependence (not shown), such that the fractions increase (decrease) with decreasing (increasing) $\beta_0$.
 
{~}\\
{\em Conclusions.}--- In this Letter, we investigated electron and ion energization in collisionless plasma turbulence in the semirelativistic regime ($m_e c^2 \lesssim T \lesssim m_i c^2$), where electrons are ultra-relativistic and ions are sub-relativistic. We used PIC simulations to perform a parameter scan in initial temperature $T_0 = \theta_{i0} m_i c^2$ that covered nearly the entire semirelativistic regime. This work thus fills a void between previous studies of turbulence in the non-relativistic regime (typically studied with reduced plasma models) and in the fully relativistic (i.e., pair plasma) regime. This study is primarily relevant for high-energy astrophysical systems with relativistic and nonthermal components, such as RIAFs.

Our results support the prevailing view that turbulent electron-ion plasmas evolve toward a non-equilibrium, ``two-temperature'' state. In particular, we find that ions reach a higher temperature than electrons in most of the explored parameter space. At a glance, this result appears to differ from non-relativistic analytical and numerical studies that find preferential electron heating at low $\beta$~\cite[e.g.,][]{howes_2010, kawazura_etal_2018}. Aside from relativistic effects, this difference can be attributed to the fact that our numerical set-up lacks an energy sink: the absence of cooling prevents low $\beta$ from being sustained for more than a few large-scale dynamical times, so simulations tend to be in the $\beta \gtrsim 1$ regime at late times, where preferential ion heating may be expected based on those previous works. A more detailed comparison of our results to non-relativistic plasmas is deferred to future work.

Our results indicate that NTPA is efficient for both species in the fully relativistic regime, but becomes inefficient for electrons when temperature is decreased through the semirelativistic regime (in contrast to ions, which continue to be efficiently accelerated). To produce hard nonthermal electron radiative signatures, astrophysical systems then require either low $\beta$ or ions with near-relativistic temperature. Cosmic ray acceleration, on the other hand, can occur even if ions are initially sub-relativistic. We caution, however, that extrapolating these conclusions regarding NTPA (and, to some extent, energy partition) to large system size is nontrivial and requires a separate scaling study \citep{zhdankin_etal_2017, zhdankin_etal_2018b}, which can perhaps be connected to MHD test-particle approaches \citep[e.g.,][]{dmitruk_etal_2004, dalena_etal_2014, lynn_etal_2014}.

This paper constitutes the first numerical investigation of plasma turbulence in the semirelativistic regime using first-principles PIC simulations. Our results, including the empirical formula for the energy partition (Eq.~\ref{eq:emp}), will be useful for modeling RIAFs and for guiding future theoretical efforts toward understanding turbulent particle energization. It is tempting to connect this empirical formula to the scaling of the turbulent fluctuations in the inertial range \citep[as in, e.g.,][]{matthaeus_etal_2016} or kinetic range, but this requires a careful analysis of the turbulence statistics and dissipation mechanisms in this regime. Hence, we leave a physical basis for this formula to future work.

\acknowledgements

The authors acknowledge support from NSF grant AST-1411879 and NASA ATP grants NNX16AB28G and NNX17AK57G. An award of computer time was provided by the Innovative and Novel Computational Impact on Theory and Experiment (INCITE) program. This research used resources of the Argonne Leadership Computing Facility, which is a DOE Office of Science User Facility supported under Contract DE-AC02-06CH11357. This work also used the Extreme Science and Engineering Discovery Environment (XSEDE), which is supported by National Science Foundation grant number ACI-1548562. This work used the XSEDE supercomputer Stampede2 at the Texas Advanced Computer Center (TACC) through allocation TG-PHY160032 \citep{xsede}.

%\bibliography{refs_all}

\begin{thebibliography}{48}%
\makeatletter
\providecommand \@ifxundefined [1]{%
 \@ifx{#1\undefined}
}%
\providecommand \@ifnum [1]{%
 \ifnum #1\expandafter \@firstoftwo
 \else \expandafter \@secondoftwo
 \fi
}%
\providecommand \@ifx [1]{%
 \ifx #1\expandafter \@firstoftwo
 \else \expandafter \@secondoftwo
 \fi
}%
\providecommand \natexlab [1]{#1}%
\providecommand \enquote  [1]{``#1''}%
\providecommand \bibnamefont  [1]{#1}%
\providecommand \bibfnamefont [1]{#1}%
\providecommand \citenamefont [1]{#1}%
\providecommand \href@noop [0]{\@secondoftwo}%
\providecommand \href [0]{\begingroup \@sanitize@url \@href}%
\providecommand \@href[1]{\@@startlink{#1}\@@href}%
\providecommand \@@href[1]{\endgroup#1\@@endlink}%
\providecommand \@sanitize@url [0]{\catcode `\\12\catcode `\$12\catcode
  `\&12\catcode `\#12\catcode `\^12\catcode `\_12\catcode `\%12\relax}%
\providecommand \@@startlink[1]{}%
\providecommand \@@endlink[0]{}%
\providecommand \url  [0]{\begingroup\@sanitize@url \@url }%
\providecommand \@url [1]{\endgroup\@href {#1}{\urlprefix }}%
\providecommand \urlprefix  [0]{URL }%
\providecommand \Eprint [0]{\href }%
\providecommand \doibase [0]{http://dx.doi.org/}%
\providecommand \selectlanguage [0]{\@gobble}%
\providecommand \bibinfo  [0]{\@secondoftwo}%
\providecommand \bibfield  [0]{\@secondoftwo}%
\providecommand \translation [1]{[#1]}%
\providecommand \BibitemOpen [0]{}%
\providecommand \bibitemStop [0]{}%
\providecommand \bibitemNoStop [0]{.\EOS\space}%
\providecommand \EOS [0]{\spacefactor3000\relax}%
\providecommand \BibitemShut  [1]{\csname bibitem#1\endcsname}%
\let\auto@bib@innerbib\@empty
%</preamble>
\bibitem [{\citenamefont {Shapiro}\ \emph {et~al.}(1976)\citenamefont
  {Shapiro}, \citenamefont {Lightman},\ and\ \citenamefont
  {Eardley}}]{shapiro_lightman_eardley_1976}%
  \BibitemOpen
  \bibfield  {author} {\bibinfo {author} {\bibfnamefont {S.}~\bibnamefont
  {Shapiro}}, \bibinfo {author} {\bibfnamefont {A.}~\bibnamefont {Lightman}}, \
  and\ \bibinfo {author} {\bibfnamefont {D.}~\bibnamefont {Eardley}},\
  }\href@noop {} {\bibfield  {journal} {\bibinfo  {journal} {Astrophysical
  Journal}\ }\textbf {\bibinfo {volume} {204}},\ \bibinfo {pages} {187}
  (\bibinfo {year} {1976})}\BibitemShut {NoStop}%
\bibitem [{\citenamefont {Ichimaru}(1977)}]{ichimaru_1977}%
  \BibitemOpen
  \bibfield  {author} {\bibinfo {author} {\bibfnamefont {S.}~\bibnamefont
  {Ichimaru}},\ }\href@noop {} {\bibfield  {journal} {\bibinfo  {journal} {The
  Astrophysical Journal}\ }\textbf {\bibinfo {volume} {214}},\ \bibinfo {pages}
  {840} (\bibinfo {year} {1977})}\BibitemShut {NoStop}%
\bibitem [{\citenamefont {Rees}\ \emph {et~al.}(1982)\citenamefont {Rees},
  \citenamefont {Begelman}, \citenamefont {Blandford},\ and\ \citenamefont
  {Phinney}}]{rees_etal_1982}%
  \BibitemOpen
  \bibfield  {author} {\bibinfo {author} {\bibfnamefont {M.}~\bibnamefont
  {Rees}}, \bibinfo {author} {\bibfnamefont {M.}~\bibnamefont {Begelman}},
  \bibinfo {author} {\bibfnamefont {R.}~\bibnamefont {Blandford}}, \ and\
  \bibinfo {author} {\bibfnamefont {E.}~\bibnamefont {Phinney}},\ }\href@noop
  {} {\bibfield  {journal} {\bibinfo  {journal} {Nature}\ }\textbf {\bibinfo
  {volume} {295}},\ \bibinfo {pages} {17} (\bibinfo {year} {1982})}\BibitemShut
  {NoStop}%
\bibitem [{\citenamefont {Narayan}\ and\ \citenamefont
  {Yi}(1995)}]{narayan_yi_1995}%
  \BibitemOpen
  \bibfield  {author} {\bibinfo {author} {\bibfnamefont {R.}~\bibnamefont
  {Narayan}}\ and\ \bibinfo {author} {\bibfnamefont {I.}~\bibnamefont {Yi}},\
  }\href@noop {} {\bibfield  {journal} {\bibinfo  {journal} {The Astrophysical
  Journal}\ }\textbf {\bibinfo {volume} {452}},\ \bibinfo {pages} {710}
  (\bibinfo {year} {1995})}\BibitemShut {NoStop}%
\bibitem [{\citenamefont {Quataert}\ and\ \citenamefont
  {Gruzinov}(1999)}]{quataert_gruzinov_1999}%
  \BibitemOpen
  \bibfield  {author} {\bibinfo {author} {\bibfnamefont {E.}~\bibnamefont
  {Quataert}}\ and\ \bibinfo {author} {\bibfnamefont {A.}~\bibnamefont
  {Gruzinov}},\ }\href@noop {} {\bibfield  {journal} {\bibinfo  {journal} {The
  Astrophysical Journal}\ }\textbf {\bibinfo {volume} {520}},\ \bibinfo {pages}
  {248} (\bibinfo {year} {1999})}\BibitemShut {NoStop}%
\bibitem [{\citenamefont {Ressler}\ \emph {et~al.}(2015)\citenamefont
  {Ressler}, \citenamefont {Tchekhovskoy}, \citenamefont {Quataert},
  \citenamefont {Chandra},\ and\ \citenamefont {Gammie}}]{ressler_etal_2015}%
  \BibitemOpen
  \bibfield  {author} {\bibinfo {author} {\bibfnamefont {S.~M.}\ \bibnamefont
  {Ressler}}, \bibinfo {author} {\bibfnamefont {A.}~\bibnamefont
  {Tchekhovskoy}}, \bibinfo {author} {\bibfnamefont {E.}~\bibnamefont
  {Quataert}}, \bibinfo {author} {\bibfnamefont {M.}~\bibnamefont {Chandra}}, \
  and\ \bibinfo {author} {\bibfnamefont {C.~F.}\ \bibnamefont {Gammie}},\
  }\href@noop {} {\bibfield  {journal} {\bibinfo  {journal} {Monthly Notices of
  the Royal Astronomical Society}\ }\textbf {\bibinfo {volume} {454}},\
  \bibinfo {pages} {1848} (\bibinfo {year} {2015})}\BibitemShut {NoStop}%
\bibitem [{\citenamefont {Foucart}\ \emph {et~al.}(2015)\citenamefont
  {Foucart}, \citenamefont {Chandra}, \citenamefont {Gammie},\ and\
  \citenamefont {Quataert}}]{foucart_etal_2015}%
  \BibitemOpen
  \bibfield  {author} {\bibinfo {author} {\bibfnamefont {F.}~\bibnamefont
  {Foucart}}, \bibinfo {author} {\bibfnamefont {M.}~\bibnamefont {Chandra}},
  \bibinfo {author} {\bibfnamefont {C.~F.}\ \bibnamefont {Gammie}}, \ and\
  \bibinfo {author} {\bibfnamefont {E.}~\bibnamefont {Quataert}},\ }\href@noop
  {} {\bibfield  {journal} {\bibinfo  {journal} {Monthly Notices of the Royal
  Astronomical Society}\ }\textbf {\bibinfo {volume} {456}},\ \bibinfo {pages}
  {1332} (\bibinfo {year} {2015})}\BibitemShut {NoStop}%
\bibitem [{\citenamefont {Ball}\ \emph
  {et~al.}(2018{\natexlab{a}})\citenamefont {Ball}, \citenamefont {{\"O}zel},
  \citenamefont {Psaltis}, \citenamefont {Chan},\ and\ \citenamefont
  {Sironi}}]{ball_etal_2018b}%
  \BibitemOpen
  \bibfield  {author} {\bibinfo {author} {\bibfnamefont {D.}~\bibnamefont
  {Ball}}, \bibinfo {author} {\bibfnamefont {F.}~\bibnamefont {{\"O}zel}},
  \bibinfo {author} {\bibfnamefont {D.}~\bibnamefont {Psaltis}}, \bibinfo
  {author} {\bibfnamefont {C.-K.}\ \bibnamefont {Chan}}, \ and\ \bibinfo
  {author} {\bibfnamefont {L.}~\bibnamefont {Sironi}},\ }\href@noop {}
  {\bibfield  {journal} {\bibinfo  {journal} {The Astrophysical Journal}\
  }\textbf {\bibinfo {volume} {853}},\ \bibinfo {pages} {184} (\bibinfo {year}
  {2018}{\natexlab{a}})}\BibitemShut {NoStop}%
\bibitem [{\citenamefont {Ryan}\ \emph {et~al.}(2018)\citenamefont {Ryan},
  \citenamefont {Ressler}, \citenamefont {Dolence}, \citenamefont {Gammie},\
  and\ \citenamefont {Quataert}}]{ryan_etal_2018}%
  \BibitemOpen
  \bibfield  {author} {\bibinfo {author} {\bibfnamefont {B.~R.}\ \bibnamefont
  {Ryan}}, \bibinfo {author} {\bibfnamefont {S.~M.}\ \bibnamefont {Ressler}},
  \bibinfo {author} {\bibfnamefont {J.~C.}\ \bibnamefont {Dolence}}, \bibinfo
  {author} {\bibfnamefont {C.~F.}\ \bibnamefont {Gammie}}, \ and\ \bibinfo
  {author} {\bibfnamefont {E.}~\bibnamefont {Quataert}},\ }\href@noop {}
  {\bibfield  {journal} {\bibinfo  {journal} {arXiv preprint arXiv:1808.01958}\
  } (\bibinfo {year} {2018})}\BibitemShut {NoStop}%
\bibitem [{\citenamefont {Howes}(2010)}]{howes_2010}%
  \BibitemOpen
  \bibfield  {author} {\bibinfo {author} {\bibfnamefont {G.~G.}\ \bibnamefont
  {Howes}},\ }\href@noop {} {\bibfield  {journal} {\bibinfo  {journal} {Monthly
  Notices of the Royal Astronomical Society: Letters}\ }\textbf {\bibinfo
  {volume} {409}},\ \bibinfo {pages} {L104} (\bibinfo {year}
  {2010})}\BibitemShut {NoStop}%
\bibitem [{\citenamefont {Matthaeus}\ \emph {et~al.}(2016)\citenamefont
  {Matthaeus}, \citenamefont {Parashar}, \citenamefont {Wan},\ and\
  \citenamefont {Wu}}]{matthaeus_etal_2016}%
  \BibitemOpen
  \bibfield  {author} {\bibinfo {author} {\bibfnamefont {W.~H.}\ \bibnamefont
  {Matthaeus}}, \bibinfo {author} {\bibfnamefont {T.~N.}\ \bibnamefont
  {Parashar}}, \bibinfo {author} {\bibfnamefont {M.}~\bibnamefont {Wan}}, \
  and\ \bibinfo {author} {\bibfnamefont {P.}~\bibnamefont {Wu}},\ }\href@noop
  {} {\bibfield  {journal} {\bibinfo  {journal} {The Astrophysical Journal
  Letters}\ }\textbf {\bibinfo {volume} {827}},\ \bibinfo {pages} {L7}
  (\bibinfo {year} {2016})}\BibitemShut {NoStop}%
\bibitem [{\citenamefont {Rowan}\ \emph {et~al.}(2017)\citenamefont {Rowan},
  \citenamefont {Sironi},\ and\ \citenamefont
  {Narayan}}]{rowan_sironi_narayan_2017}%
  \BibitemOpen
  \bibfield  {author} {\bibinfo {author} {\bibfnamefont {M.~E.}\ \bibnamefont
  {Rowan}}, \bibinfo {author} {\bibfnamefont {L.}~\bibnamefont {Sironi}}, \
  and\ \bibinfo {author} {\bibfnamefont {R.}~\bibnamefont {Narayan}},\
  }\href@noop {} {\bibfield  {journal} {\bibinfo  {journal} {The Astrophysical
  Journal}\ }\textbf {\bibinfo {volume} {850}},\ \bibinfo {pages} {29}
  (\bibinfo {year} {2017})}\BibitemShut {NoStop}%
\bibitem [{\citenamefont {Werner}\ \emph {et~al.}(2018)\citenamefont {Werner},
  \citenamefont {Uzdensky}, \citenamefont {Begelman}, \citenamefont {Cerutti},\
  and\ \citenamefont {Nalewajko}}]{werner_etal_2018}%
  \BibitemOpen
  \bibfield  {author} {\bibinfo {author} {\bibfnamefont {G.}~\bibnamefont
  {Werner}}, \bibinfo {author} {\bibfnamefont {D.}~\bibnamefont {Uzdensky}},
  \bibinfo {author} {\bibfnamefont {M.}~\bibnamefont {Begelman}}, \bibinfo
  {author} {\bibfnamefont {B.}~\bibnamefont {Cerutti}}, \ and\ \bibinfo
  {author} {\bibfnamefont {K.}~\bibnamefont {Nalewajko}},\ }\href@noop {}
  {\bibfield  {journal} {\bibinfo  {journal} {Monthly Notices of the Royal
  Astronomical Society}\ }\textbf {\bibinfo {volume} {473}},\ \bibinfo {pages}
  {4840} (\bibinfo {year} {2018})}\BibitemShut {NoStop}%
\bibitem [{\citenamefont {{Kawazura}}\ \emph {et~al.}(2018)\citenamefont
  {{Kawazura}}, \citenamefont {{Barnes}},\ and\ \citenamefont
  {{Schekochihin}}}]{kawazura_etal_2018}%
  \BibitemOpen
  \bibfield  {author} {\bibinfo {author} {\bibfnamefont {Y.}~\bibnamefont
  {{Kawazura}}}, \bibinfo {author} {\bibfnamefont {M.}~\bibnamefont
  {{Barnes}}}, \ and\ \bibinfo {author} {\bibfnamefont {A.~A.}\ \bibnamefont
  {{Schekochihin}}},\ }\href@noop {} {\bibfield  {journal} {\bibinfo  {journal}
  {ArXiv e-prints}\ } (\bibinfo {year} {2018})},\ \Eprint
  {http://arxiv.org/abs/1807.07702} {arXiv:1807.07702 [physics.plasm-ph]}
  \BibitemShut {NoStop}%
\bibitem [{\citenamefont {Begelman}\ and\ \citenamefont
  {Chiueh}(1988)}]{begelman_chiueh_1988}%
  \BibitemOpen
  \bibfield  {author} {\bibinfo {author} {\bibfnamefont {M.~C.}\ \bibnamefont
  {Begelman}}\ and\ \bibinfo {author} {\bibfnamefont {T.}~\bibnamefont
  {Chiueh}},\ }\href@noop {} {\bibfield  {journal} {\bibinfo  {journal} {The
  Astrophysical Journal}\ }\textbf {\bibinfo {volume} {332}},\ \bibinfo {pages}
  {872} (\bibinfo {year} {1988})}\BibitemShut {NoStop}%
\bibitem [{\citenamefont {Sironi}\ and\ \citenamefont
  {Narayan}(2015)}]{sironi_narayan_2015}%
  \BibitemOpen
  \bibfield  {author} {\bibinfo {author} {\bibfnamefont {L.}~\bibnamefont
  {Sironi}}\ and\ \bibinfo {author} {\bibfnamefont {R.}~\bibnamefont
  {Narayan}},\ }\href@noop {} {\bibfield  {journal} {\bibinfo  {journal} {The
  Astrophysical Journal}\ }\textbf {\bibinfo {volume} {800}},\ \bibinfo {pages}
  {88} (\bibinfo {year} {2015})}\BibitemShut {NoStop}%
\bibitem [{\citenamefont {Yuan}\ \emph {et~al.}(2003)\citenamefont {Yuan},
  \citenamefont {Quataert},\ and\ \citenamefont {Narayan}}]{yuan_etal_2003}%
  \BibitemOpen
  \bibfield  {author} {\bibinfo {author} {\bibfnamefont {F.}~\bibnamefont
  {Yuan}}, \bibinfo {author} {\bibfnamefont {E.}~\bibnamefont {Quataert}}, \
  and\ \bibinfo {author} {\bibfnamefont {R.}~\bibnamefont {Narayan}},\
  }\href@noop {} {\bibfield  {journal} {\bibinfo  {journal} {The Astrophysical
  Journal}\ }\textbf {\bibinfo {volume} {598}},\ \bibinfo {pages} {301}
  (\bibinfo {year} {2003})}\BibitemShut {NoStop}%
\bibitem [{\citenamefont {Remillard}\ and\ \citenamefont
  {McClintock}(2006)}]{remillard_mcclintock_2006}%
  \BibitemOpen
  \bibfield  {author} {\bibinfo {author} {\bibfnamefont {R.~A.}\ \bibnamefont
  {Remillard}}\ and\ \bibinfo {author} {\bibfnamefont {J.~E.}\ \bibnamefont
  {McClintock}},\ }\href@noop {} {\bibfield  {journal} {\bibinfo  {journal}
  {Annu. Rev. Astron. Astrophys.}\ }\textbf {\bibinfo {volume} {44}},\ \bibinfo
  {pages} {49} (\bibinfo {year} {2006})}\BibitemShut {NoStop}%
\bibitem [{\citenamefont {Yuan}\ and\ \citenamefont
  {Narayan}(2014)}]{yuan_narayan_2014}%
  \BibitemOpen
  \bibfield  {author} {\bibinfo {author} {\bibfnamefont {F.}~\bibnamefont
  {Yuan}}\ and\ \bibinfo {author} {\bibfnamefont {R.}~\bibnamefont {Narayan}},\
  }\href@noop {} {\bibfield  {journal} {\bibinfo  {journal} {Annual Review of
  Astronomy and Astrophysics}\ }\textbf {\bibinfo {volume} {52}},\ \bibinfo
  {pages} {529} (\bibinfo {year} {2014})}\BibitemShut {NoStop}%
\bibitem [{\citenamefont {Balbus}\ and\ \citenamefont
  {Hawley}(1991)}]{balbus_hawley_1991}%
  \BibitemOpen
  \bibfield  {author} {\bibinfo {author} {\bibfnamefont {S.~A.}\ \bibnamefont
  {Balbus}}\ and\ \bibinfo {author} {\bibfnamefont {J.~F.}\ \bibnamefont
  {Hawley}},\ }\href@noop {} {\bibfield  {journal} {\bibinfo  {journal} {The
  Astrophysical Journal}\ }\textbf {\bibinfo {volume} {376}},\ \bibinfo {pages}
  {214} (\bibinfo {year} {1991})}\BibitemShut {NoStop}%
\bibitem [{\citenamefont {Riquelme}\ \emph {et~al.}(2012)\citenamefont
  {Riquelme}, \citenamefont {Quataert}, \citenamefont {Sharma},\ and\
  \citenamefont {Spitkovsky}}]{riquelme_etal_2012}%
  \BibitemOpen
  \bibfield  {author} {\bibinfo {author} {\bibfnamefont {M.~A.}\ \bibnamefont
  {Riquelme}}, \bibinfo {author} {\bibfnamefont {E.}~\bibnamefont {Quataert}},
  \bibinfo {author} {\bibfnamefont {P.}~\bibnamefont {Sharma}}, \ and\ \bibinfo
  {author} {\bibfnamefont {A.}~\bibnamefont {Spitkovsky}},\ }\href@noop {}
  {\bibfield  {journal} {\bibinfo  {journal} {The Astrophysical Journal}\
  }\textbf {\bibinfo {volume} {755}},\ \bibinfo {pages} {50} (\bibinfo {year}
  {2012})}\BibitemShut {NoStop}%
\bibitem [{\citenamefont {Hoshino}(2013)}]{hoshino_2013}%
  \BibitemOpen
  \bibfield  {author} {\bibinfo {author} {\bibfnamefont {M.}~\bibnamefont
  {Hoshino}},\ }\href@noop {} {\bibfield  {journal} {\bibinfo  {journal} {The
  Astrophysical Journal}\ }\textbf {\bibinfo {volume} {773}},\ \bibinfo {pages}
  {118} (\bibinfo {year} {2013})}\BibitemShut {NoStop}%
\bibitem [{\citenamefont {Hoshino}(2015)}]{hoshino_2015}%
  \BibitemOpen
  \bibfield  {author} {\bibinfo {author} {\bibfnamefont {M.}~\bibnamefont
  {Hoshino}},\ }\href@noop {} {\bibfield  {journal} {\bibinfo  {journal}
  {Physical Review Letters}\ }\textbf {\bibinfo {volume} {114}},\ \bibinfo
  {pages} {061101} (\bibinfo {year} {2015})}\BibitemShut {NoStop}%
\bibitem [{\citenamefont {Kunz}\ \emph {et~al.}(2016)\citenamefont {Kunz},
  \citenamefont {Stone},\ and\ \citenamefont {Quataert}}]{kunz_etal_2016}%
  \BibitemOpen
  \bibfield  {author} {\bibinfo {author} {\bibfnamefont {M.~W.}\ \bibnamefont
  {Kunz}}, \bibinfo {author} {\bibfnamefont {J.~M.}\ \bibnamefont {Stone}}, \
  and\ \bibinfo {author} {\bibfnamefont {E.}~\bibnamefont {Quataert}},\
  }\href@noop {} {\bibfield  {journal} {\bibinfo  {journal} {Physical Review
  Letters}\ }\textbf {\bibinfo {volume} {117}},\ \bibinfo {pages} {235101}
  (\bibinfo {year} {2016})}\BibitemShut {NoStop}%
\bibitem [{\citenamefont {Inchingolo}\ \emph {et~al.}(2018)\citenamefont
  {Inchingolo}, \citenamefont {Grismayer}, \citenamefont {Loureiro},
  \citenamefont {Fonseca},\ and\ \citenamefont {Silva}}]{inchingolo_etal_2018}%
  \BibitemOpen
  \bibfield  {author} {\bibinfo {author} {\bibfnamefont {G.}~\bibnamefont
  {Inchingolo}}, \bibinfo {author} {\bibfnamefont {T.}~\bibnamefont
  {Grismayer}}, \bibinfo {author} {\bibfnamefont {N.~F.}\ \bibnamefont
  {Loureiro}}, \bibinfo {author} {\bibfnamefont {R.~A.}\ \bibnamefont
  {Fonseca}}, \ and\ \bibinfo {author} {\bibfnamefont {L.~O.}\ \bibnamefont
  {Silva}},\ }\href@noop {} {\bibfield  {journal} {\bibinfo  {journal} {The
  Astrophysical Journal}\ }\textbf {\bibinfo {volume} {859}},\ \bibinfo {pages}
  {149} (\bibinfo {year} {2018})}\BibitemShut {NoStop}%
\bibitem [{\citenamefont {Kimura}\ \emph {et~al.}(2016)\citenamefont {Kimura},
  \citenamefont {Toma}, \citenamefont {Suzuki},\ and\ \citenamefont
  {Inutsuka}}]{kimura_etal_2016}%
  \BibitemOpen
  \bibfield  {author} {\bibinfo {author} {\bibfnamefont {S.~S.}\ \bibnamefont
  {Kimura}}, \bibinfo {author} {\bibfnamefont {K.}~\bibnamefont {Toma}},
  \bibinfo {author} {\bibfnamefont {T.~K.}\ \bibnamefont {Suzuki}}, \ and\
  \bibinfo {author} {\bibfnamefont {S.-i.}\ \bibnamefont {Inutsuka}},\
  }\href@noop {} {\bibfield  {journal} {\bibinfo  {journal} {The Astrophysical
  Journal}\ }\textbf {\bibinfo {volume} {822}},\ \bibinfo {pages} {88}
  (\bibinfo {year} {2016})}\BibitemShut {NoStop}%
\bibitem [{\citenamefont {Guo}\ \emph {et~al.}(2016)\citenamefont {Guo},
  \citenamefont {Li}, \citenamefont {Li}, \citenamefont {Daughton},
  \citenamefont {Zhang}, \citenamefont {Lloyd-Ronning}, \citenamefont {Liu},
  \citenamefont {Zhang},\ and\ \citenamefont {Deng}}]{guo_etal_2016}%
  \BibitemOpen
  \bibfield  {author} {\bibinfo {author} {\bibfnamefont {F.}~\bibnamefont
  {Guo}}, \bibinfo {author} {\bibfnamefont {X.}~\bibnamefont {Li}}, \bibinfo
  {author} {\bibfnamefont {H.}~\bibnamefont {Li}}, \bibinfo {author}
  {\bibfnamefont {W.}~\bibnamefont {Daughton}}, \bibinfo {author}
  {\bibfnamefont {B.}~\bibnamefont {Zhang}}, \bibinfo {author} {\bibfnamefont
  {N.}~\bibnamefont {Lloyd-Ronning}}, \bibinfo {author} {\bibfnamefont {Y.-H.}\
  \bibnamefont {Liu}}, \bibinfo {author} {\bibfnamefont {H.}~\bibnamefont
  {Zhang}}, \ and\ \bibinfo {author} {\bibfnamefont {W.}~\bibnamefont {Deng}},\
  }\href@noop {} {\bibfield  {journal} {\bibinfo  {journal} {The Astrophysical
  Journal Letters}\ }\textbf {\bibinfo {volume} {818}},\ \bibinfo {pages} {L9}
  (\bibinfo {year} {2016})}\BibitemShut {NoStop}%
\bibitem [{\citenamefont {Ball}\ \emph
  {et~al.}(2018{\natexlab{b}})\citenamefont {Ball}, \citenamefont {Sironi},\
  and\ \citenamefont {{\"O}zel}}]{ball_etal_2018}%
  \BibitemOpen
  \bibfield  {author} {\bibinfo {author} {\bibfnamefont {D.}~\bibnamefont
  {Ball}}, \bibinfo {author} {\bibfnamefont {L.}~\bibnamefont {Sironi}}, \ and\
  \bibinfo {author} {\bibfnamefont {F.}~\bibnamefont {{\"O}zel}},\ }\href@noop
  {} {\bibfield  {journal} {\bibinfo  {journal} {arXiv preprint
  arXiv:1803.05556}\ } (\bibinfo {year} {2018}{\natexlab{b}})}\BibitemShut
  {NoStop}%
\bibitem [{\citenamefont {Zhdankin}\ \emph {et~al.}(2018)\citenamefont
  {Zhdankin}, \citenamefont {Uzdensky}, \citenamefont {Werner},\ and\
  \citenamefont {Begelman}}]{zhdankin_etal_2018a}%
  \BibitemOpen
  \bibfield  {author} {\bibinfo {author} {\bibfnamefont {V.}~\bibnamefont
  {Zhdankin}}, \bibinfo {author} {\bibfnamefont {D.~A.}\ \bibnamefont
  {Uzdensky}}, \bibinfo {author} {\bibfnamefont {G.~R.}\ \bibnamefont
  {Werner}}, \ and\ \bibinfo {author} {\bibfnamefont {M.~C.}\ \bibnamefont
  {Begelman}},\ }\href@noop {} {\bibfield  {journal} {\bibinfo  {journal}
  {Monthly Notices of the Royal Astronomical Society}\ }\textbf {\bibinfo
  {volume} {474}},\ \bibinfo {pages} {2514} (\bibinfo {year}
  {2018})}\BibitemShut {NoStop}%
\bibitem [{\citenamefont {Cerutti}\ \emph {et~al.}(2013)\citenamefont
  {Cerutti}, \citenamefont {Werner}, \citenamefont {Uzdensky},\ and\
  \citenamefont {Begelman}}]{cerutti_etal_2013}%
  \BibitemOpen
  \bibfield  {author} {\bibinfo {author} {\bibfnamefont {B.}~\bibnamefont
  {Cerutti}}, \bibinfo {author} {\bibfnamefont {G.~R.}\ \bibnamefont {Werner}},
  \bibinfo {author} {\bibfnamefont {D.~A.}\ \bibnamefont {Uzdensky}}, \ and\
  \bibinfo {author} {\bibfnamefont {M.~C.}\ \bibnamefont {Begelman}},\
  }\href@noop {} {\bibfield  {journal} {\bibinfo  {journal} {The Astrophysical
  Journal}\ }\textbf {\bibinfo {volume} {770}},\ \bibinfo {pages} {147}
  (\bibinfo {year} {2013})}\BibitemShut {NoStop}%
\bibitem [{\citenamefont {Esirkepov}(2001)}]{esirkepov_2001}%
  \BibitemOpen
  \bibfield  {author} {\bibinfo {author} {\bibfnamefont {T.~Z.}\ \bibnamefont
  {Esirkepov}},\ }\href@noop {} {\bibfield  {journal} {\bibinfo  {journal}
  {Computer Physics Communications}\ }\textbf {\bibinfo {volume} {135}},\
  \bibinfo {pages} {144} (\bibinfo {year} {2001})}\BibitemShut {NoStop}%
\bibitem [{\citenamefont {TenBarge}\ \emph {et~al.}(2014)\citenamefont
  {TenBarge}, \citenamefont {Howes}, \citenamefont {Dorland},\ and\
  \citenamefont {Hammett}}]{tenbarge_etal_2014}%
  \BibitemOpen
  \bibfield  {author} {\bibinfo {author} {\bibfnamefont {J.}~\bibnamefont
  {TenBarge}}, \bibinfo {author} {\bibfnamefont {G.~G.}\ \bibnamefont {Howes}},
  \bibinfo {author} {\bibfnamefont {W.}~\bibnamefont {Dorland}}, \ and\
  \bibinfo {author} {\bibfnamefont {G.~W.}\ \bibnamefont {Hammett}},\
  }\href@noop {} {\bibfield  {journal} {\bibinfo  {journal} {Computer Physics
  Communications}\ }\textbf {\bibinfo {volume} {185}},\ \bibinfo {pages} {578}
  (\bibinfo {year} {2014})}\BibitemShut {NoStop}%
\bibitem [{\citenamefont {{Zhdankin}}\ \emph {et~al.}(2018)\citenamefont
  {{Zhdankin}}, \citenamefont {{Uzdensky}}, \citenamefont {{Werner}},\ and\
  \citenamefont {{Begelman}}}]{zhdankin_etal_2018b}%
  \BibitemOpen
  \bibfield  {author} {\bibinfo {author} {\bibfnamefont {V.}~\bibnamefont
  {{Zhdankin}}}, \bibinfo {author} {\bibfnamefont {D.~A.}\ \bibnamefont
  {{Uzdensky}}}, \bibinfo {author} {\bibfnamefont {G.~R.}\ \bibnamefont
  {{Werner}}}, \ and\ \bibinfo {author} {\bibfnamefont {M.~C.}\ \bibnamefont
  {{Begelman}}},\ }\href@noop {} {\bibfield  {journal} {\bibinfo  {journal}
  {The Astrophysical Journal Letters}\ }\textbf {\bibinfo {volume} {867}},\
  \bibinfo {eid} {L18} (\bibinfo {year} {2018})}\BibitemShut {NoStop}%
\bibitem [{\citenamefont {Goldreich}\ and\ \citenamefont
  {Sridhar}(1995)}]{goldreich_sridhar_1995}%
  \BibitemOpen
  \bibfield  {author} {\bibinfo {author} {\bibfnamefont {P.}~\bibnamefont
  {Goldreich}}\ and\ \bibinfo {author} {\bibfnamefont {S.}~\bibnamefont
  {Sridhar}},\ }\href@noop {} {\bibfield  {journal} {\bibinfo  {journal} {The
  Astrophysical Journal}\ }\textbf {\bibinfo {volume} {438}},\ \bibinfo {pages}
  {763} (\bibinfo {year} {1995})}\BibitemShut {NoStop}%
\bibitem [{\citenamefont {Alexandrova}\ \emph {et~al.}(2009)\citenamefont
  {Alexandrova}, \citenamefont {Saur}, \citenamefont {Lacombe}, \citenamefont
  {Mangeney}, \citenamefont {Mitchell}, \citenamefont {Schwartz},\ and\
  \citenamefont {Robert}}]{alexandrova_etal_2009}%
  \BibitemOpen
  \bibfield  {author} {\bibinfo {author} {\bibfnamefont {O.}~\bibnamefont
  {Alexandrova}}, \bibinfo {author} {\bibfnamefont {J.}~\bibnamefont {Saur}},
  \bibinfo {author} {\bibfnamefont {C.}~\bibnamefont {Lacombe}}, \bibinfo
  {author} {\bibfnamefont {A.}~\bibnamefont {Mangeney}}, \bibinfo {author}
  {\bibfnamefont {J.}~\bibnamefont {Mitchell}}, \bibinfo {author}
  {\bibfnamefont {S.~J.}\ \bibnamefont {Schwartz}}, \ and\ \bibinfo {author}
  {\bibfnamefont {P.}~\bibnamefont {Robert}},\ }\href@noop {} {\bibfield
  {journal} {\bibinfo  {journal} {Physical Review Letters}\ }\textbf {\bibinfo
  {volume} {103}},\ \bibinfo {pages} {165003} (\bibinfo {year}
  {2009})}\BibitemShut {NoStop}%
\bibitem [{\citenamefont {Sahraoui}\ \emph {et~al.}(2009)\citenamefont
  {Sahraoui}, \citenamefont {Goldstein}, \citenamefont {Robert},\ and\
  \citenamefont {Khotyaintsev}}]{sahraoui_etal_2009}%
  \BibitemOpen
  \bibfield  {author} {\bibinfo {author} {\bibfnamefont {F.}~\bibnamefont
  {Sahraoui}}, \bibinfo {author} {\bibfnamefont {M.}~\bibnamefont {Goldstein}},
  \bibinfo {author} {\bibfnamefont {P.}~\bibnamefont {Robert}}, \ and\ \bibinfo
  {author} {\bibfnamefont {Y.~V.}\ \bibnamefont {Khotyaintsev}},\ }\href@noop
  {} {\bibfield  {journal} {\bibinfo  {journal} {Physical Review Letters}\
  }\textbf {\bibinfo {volume} {102}},\ \bibinfo {pages} {231102} (\bibinfo
  {year} {2009})}\BibitemShut {NoStop}%
\bibitem [{\citenamefont {Kiyani}\ \emph {et~al.}(2009)\citenamefont {Kiyani},
  \citenamefont {Chapman}, \citenamefont {Khotyaintsev}, \citenamefont
  {Dunlop},\ and\ \citenamefont {Sahraoui}}]{kiyani_etal_2009}%
  \BibitemOpen
  \bibfield  {author} {\bibinfo {author} {\bibfnamefont {K.}~\bibnamefont
  {Kiyani}}, \bibinfo {author} {\bibfnamefont {S.}~\bibnamefont {Chapman}},
  \bibinfo {author} {\bibfnamefont {Y.~V.}\ \bibnamefont {Khotyaintsev}},
  \bibinfo {author} {\bibfnamefont {M.}~\bibnamefont {Dunlop}}, \ and\ \bibinfo
  {author} {\bibfnamefont {F.}~\bibnamefont {Sahraoui}},\ }\href@noop {}
  {\bibfield  {journal} {\bibinfo  {journal} {Physical Review Letters}\
  }\textbf {\bibinfo {volume} {103}},\ \bibinfo {pages} {075006} (\bibinfo
  {year} {2009})}\BibitemShut {NoStop}%
\bibitem [{\citenamefont {Alexandrova}\ \emph {et~al.}(2012)\citenamefont
  {Alexandrova}, \citenamefont {Lacombe}, \citenamefont {Mangeney},
  \citenamefont {Grappin},\ and\ \citenamefont
  {Maksimovic}}]{alexandrova_etal_2012}%
  \BibitemOpen
  \bibfield  {author} {\bibinfo {author} {\bibfnamefont {O.}~\bibnamefont
  {Alexandrova}}, \bibinfo {author} {\bibfnamefont {C.}~\bibnamefont
  {Lacombe}}, \bibinfo {author} {\bibfnamefont {A.}~\bibnamefont {Mangeney}},
  \bibinfo {author} {\bibfnamefont {R.}~\bibnamefont {Grappin}}, \ and\
  \bibinfo {author} {\bibfnamefont {M.}~\bibnamefont {Maksimovic}},\
  }\href@noop {} {\bibfield  {journal} {\bibinfo  {journal} {The Astrophysical
  Journal}\ }\textbf {\bibinfo {volume} {760}},\ \bibinfo {pages} {121}
  (\bibinfo {year} {2012})}\BibitemShut {NoStop}%
\bibitem [{\citenamefont {Kiyani}\ \emph {et~al.}(2015)\citenamefont {Kiyani},
  \citenamefont {Osman},\ and\ \citenamefont {Chapman}}]{kiyani_etal_2015}%
  \BibitemOpen
  \bibfield  {author} {\bibinfo {author} {\bibfnamefont {K.~H.}\ \bibnamefont
  {Kiyani}}, \bibinfo {author} {\bibfnamefont {K.~T.}\ \bibnamefont {Osman}}, \
  and\ \bibinfo {author} {\bibfnamefont {S.~C.}\ \bibnamefont {Chapman}},\
  }\href@noop {} {\bibfield  {journal} {\bibinfo  {journal} {Philosophical
  Transactions of the Royal Society of London A: Mathematical, Physical and
  Engineering Sciences}\ }\textbf {\bibinfo {volume} {373}},\ \bibinfo {pages}
  {20140155} (\bibinfo {year} {2015})}\BibitemShut {NoStop}%
\bibitem [{\citenamefont {Boldyrev}\ and\ \citenamefont
  {Perez}(2012)}]{boldyrev_perez_2012}%
  \BibitemOpen
  \bibfield  {author} {\bibinfo {author} {\bibfnamefont {S.}~\bibnamefont
  {Boldyrev}}\ and\ \bibinfo {author} {\bibfnamefont {J.~C.}\ \bibnamefont
  {Perez}},\ }\href@noop {} {\bibfield  {journal} {\bibinfo  {journal} {The
  Astrophysical Journal Letters}\ }\textbf {\bibinfo {volume} {758}},\ \bibinfo
  {pages} {L44} (\bibinfo {year} {2012})}\BibitemShut {NoStop}%
\bibitem [{\citenamefont {Told}\ \emph {et~al.}(2015)\citenamefont {Told},
  \citenamefont {Jenko}, \citenamefont {TenBarge}, \citenamefont {Howes},\ and\
  \citenamefont {Hammett}}]{told_etal_2015}%
  \BibitemOpen
  \bibfield  {author} {\bibinfo {author} {\bibfnamefont {D.}~\bibnamefont
  {Told}}, \bibinfo {author} {\bibfnamefont {F.}~\bibnamefont {Jenko}},
  \bibinfo {author} {\bibfnamefont {J.}~\bibnamefont {TenBarge}}, \bibinfo
  {author} {\bibfnamefont {G.}~\bibnamefont {Howes}}, \ and\ \bibinfo {author}
  {\bibfnamefont {G.}~\bibnamefont {Hammett}},\ }\href@noop {} {\bibfield
  {journal} {\bibinfo  {journal} {Physical Review Letters}\ }\textbf {\bibinfo
  {volume} {115}},\ \bibinfo {pages} {025003} (\bibinfo {year}
  {2015})}\BibitemShut {NoStop}%
\bibitem [{\citenamefont {Cerri}\ \emph {et~al.}(2017)\citenamefont {Cerri},
  \citenamefont {Servidio},\ and\ \citenamefont {Califano}}]{cerri_etal_2017}%
  \BibitemOpen
  \bibfield  {author} {\bibinfo {author} {\bibfnamefont {S.~S.}\ \bibnamefont
  {Cerri}}, \bibinfo {author} {\bibfnamefont {S.}~\bibnamefont {Servidio}}, \
  and\ \bibinfo {author} {\bibfnamefont {F.}~\bibnamefont {Califano}},\
  }\href@noop {} {\bibfield  {journal} {\bibinfo  {journal} {The Astrophysical
  Journal Letters}\ }\textbf {\bibinfo {volume} {846}},\ \bibinfo {pages} {L18}
  (\bibinfo {year} {2017})}\BibitemShut {NoStop}%
\bibitem [{\citenamefont {Gro{\v{s}}elj}\ \emph {et~al.}(2018)\citenamefont
  {Gro{\v{s}}elj}, \citenamefont {Mallet}, \citenamefont {Loureiro},\ and\
  \citenamefont {Jenko}}]{groselj_etal_2018}%
  \BibitemOpen
  \bibfield  {author} {\bibinfo {author} {\bibfnamefont {D.}~\bibnamefont
  {Gro{\v{s}}elj}}, \bibinfo {author} {\bibfnamefont {A.}~\bibnamefont
  {Mallet}}, \bibinfo {author} {\bibfnamefont {N.~F.}\ \bibnamefont
  {Loureiro}}, \ and\ \bibinfo {author} {\bibfnamefont {F.}~\bibnamefont
  {Jenko}},\ }\href@noop {} {\bibfield  {journal} {\bibinfo  {journal}
  {Physical Review Letters}\ }\textbf {\bibinfo {volume} {120}},\ \bibinfo
  {pages} {105101} (\bibinfo {year} {2018})}\BibitemShut {NoStop}%
\bibitem [{\citenamefont {Zhdankin}\ \emph {et~al.}(2017)\citenamefont
  {Zhdankin}, \citenamefont {Werner}, \citenamefont {Uzdensky},\ and\
  \citenamefont {Begelman}}]{zhdankin_etal_2017}%
  \BibitemOpen
  \bibfield  {author} {\bibinfo {author} {\bibfnamefont {V.}~\bibnamefont
  {Zhdankin}}, \bibinfo {author} {\bibfnamefont {G.~R.}\ \bibnamefont
  {Werner}}, \bibinfo {author} {\bibfnamefont {D.~A.}\ \bibnamefont
  {Uzdensky}}, \ and\ \bibinfo {author} {\bibfnamefont {M.~C.}\ \bibnamefont
  {Begelman}},\ }\href@noop {} {\bibfield  {journal} {\bibinfo  {journal}
  {Physical Review Letters}\ }\textbf {\bibinfo {volume} {118}},\ \bibinfo
  {pages} {055103} (\bibinfo {year} {2017})}\BibitemShut {NoStop}%
\bibitem [{\citenamefont {Dmitruk}\ \emph {et~al.}(2004)\citenamefont
  {Dmitruk}, \citenamefont {Matthaeus},\ and\ \citenamefont
  {Seenu}}]{dmitruk_etal_2004}%
  \BibitemOpen
  \bibfield  {author} {\bibinfo {author} {\bibfnamefont {P.}~\bibnamefont
  {Dmitruk}}, \bibinfo {author} {\bibfnamefont {W.}~\bibnamefont {Matthaeus}},
  \ and\ \bibinfo {author} {\bibfnamefont {N.}~\bibnamefont {Seenu}},\
  }\href@noop {} {\bibfield  {journal} {\bibinfo  {journal} {The Astrophysical
  Journal}\ }\textbf {\bibinfo {volume} {617}},\ \bibinfo {pages} {667}
  (\bibinfo {year} {2004})}\BibitemShut {NoStop}%
\bibitem [{\citenamefont {Dalena}\ \emph {et~al.}(2014)\citenamefont {Dalena},
  \citenamefont {Rappazzo}, \citenamefont {Dmitruk}, \citenamefont {Greco},\
  and\ \citenamefont {Matthaeus}}]{dalena_etal_2014}%
  \BibitemOpen
  \bibfield  {author} {\bibinfo {author} {\bibfnamefont {S.}~\bibnamefont
  {Dalena}}, \bibinfo {author} {\bibfnamefont {A.}~\bibnamefont {Rappazzo}},
  \bibinfo {author} {\bibfnamefont {P.}~\bibnamefont {Dmitruk}}, \bibinfo
  {author} {\bibfnamefont {A.}~\bibnamefont {Greco}}, \ and\ \bibinfo {author}
  {\bibfnamefont {W.}~\bibnamefont {Matthaeus}},\ }\href@noop {} {\bibfield
  {journal} {\bibinfo  {journal} {The Astrophysical Journal}\ }\textbf
  {\bibinfo {volume} {783}},\ \bibinfo {pages} {143} (\bibinfo {year}
  {2014})}\BibitemShut {NoStop}%
\bibitem [{\citenamefont {Lynn}\ \emph {et~al.}(2014)\citenamefont {Lynn},
  \citenamefont {Quataert}, \citenamefont {Chandran},\ and\ \citenamefont
  {Parrish}}]{lynn_etal_2014}%
  \BibitemOpen
  \bibfield  {author} {\bibinfo {author} {\bibfnamefont {J.~W.}\ \bibnamefont
  {Lynn}}, \bibinfo {author} {\bibfnamefont {E.}~\bibnamefont {Quataert}},
  \bibinfo {author} {\bibfnamefont {B.~D.}\ \bibnamefont {Chandran}}, \ and\
  \bibinfo {author} {\bibfnamefont {I.~J.}\ \bibnamefont {Parrish}},\
  }\href@noop {} {\bibfield  {journal} {\bibinfo  {journal} {The Astrophysical
  Journal}\ }\textbf {\bibinfo {volume} {791}},\ \bibinfo {pages} {71}
  (\bibinfo {year} {2014})}\BibitemShut {NoStop}%
\bibitem [{\citenamefont {Towns}\ \emph {et~al.}(2014)\citenamefont {Towns},
  \citenamefont {Cockerill}, \citenamefont {Dahan}, \citenamefont {Foster},
  \citenamefont {Gaither}, \citenamefont {Grimshaw}, \citenamefont {Hazlewood},
  \citenamefont {Lathrop}, \citenamefont {Lifka}, \citenamefont {Peterson},
  \citenamefont {Roskies}, \citenamefont {Scott},\ and\ \citenamefont
  {Wilkins-Diehr}}]{xsede}%
  \BibitemOpen
  \bibfield  {author} {\bibinfo {author} {\bibfnamefont {J.}~\bibnamefont
  {Towns}}, \bibinfo {author} {\bibfnamefont {T.}~\bibnamefont {Cockerill}},
  \bibinfo {author} {\bibfnamefont {M.}~\bibnamefont {Dahan}}, \bibinfo
  {author} {\bibfnamefont {I.}~\bibnamefont {Foster}}, \bibinfo {author}
  {\bibfnamefont {K.}~\bibnamefont {Gaither}}, \bibinfo {author} {\bibfnamefont
  {A.}~\bibnamefont {Grimshaw}}, \bibinfo {author} {\bibfnamefont
  {V.}~\bibnamefont {Hazlewood}}, \bibinfo {author} {\bibfnamefont
  {S.}~\bibnamefont {Lathrop}}, \bibinfo {author} {\bibfnamefont
  {D.}~\bibnamefont {Lifka}}, \bibinfo {author} {\bibfnamefont {G.~D.}\
  \bibnamefont {Peterson}}, \bibinfo {author} {\bibfnamefont {R.}~\bibnamefont
  {Roskies}}, \bibinfo {author} {\bibfnamefont {J.~R.}\ \bibnamefont {Scott}},
  \ and\ \bibinfo {author} {\bibfnamefont {N.}~\bibnamefont {Wilkins-Diehr}},\
  }\href {\doibase 10.1109/MCSE.2014.80} {\bibfield  {journal} {\bibinfo
  {journal} {Computing in Science \& Engineering}\ }\textbf {\bibinfo {volume}
  {16}},\ \bibinfo {pages} {62} (\bibinfo {year} {2014})}\BibitemShut {NoStop}%
\end{thebibliography}

%

\end{document}